\documentclass[prd,twocolumn,showpacs]{revtex4-1}
\usepackage{amsfonts}
\usepackage[english]{babel}
\usepackage{amsmath,amssymb,amsthm} 
\usepackage{amscd}
\usepackage{graphics}
\usepackage{latexsym}
\usepackage[colorlinks,citecolor=blue,urlcolor=blue,linkcolor=blue]{hyperref}
\newlength{\altura}
\newcommand{\old}[1]{\settoheight{\altura}{$#1$}{\stackrel{{\rm o}}{#1}}\rule[0pt]{0pt}{\altura}}

\begin{document}
\title{Teleparallel Darkness.}
\author{M. Hermida de La Rica}
\affiliation{Universidad Polit\'ecnica de Madrid.\\
University of Ngozi.}

\begin{abstract}
First a review of Teleparallel theory is done with special emphasis in
the derivation of conservation equations within this theory and in
particular of energy-momentum conservation. Given that we are allowed
to speak about the existence of negative energy, the question is that
in its interaction with matter, we need not have matter conservation:
It is only the sum of both which should remain constant. This does not
only leads to an accelerated expansion without the need of a
cosmological constant, but it may also contribute to explain the
origin of dark matter, and poses questions about the origin of
inflation at earlier times. The prediction of the proposed model can
be qualitatively compared to recent results of Cosmic Microwave
Background (CMB) analysis.
\end{abstract}
\maketitle

\section{Conservation equations}
We follow most of the conventions and notations of references
\cite{Andrade2000} and \cite{Andrade1997}. We also adopt the
$(+,-,-,-)$ Lorentzian metric convention. Let us also clarify that
greek indices are used for the coordinate holonomic quantities and
latin indices are used for non-holonomic ones.

One way of viewing the kind of Riemannian spaces in which Teleparallel
theories are formulated, is by taking as departure point the hypothesis
that physics somehow establishes a canonical, smooth, path-independent
isomorphism between the tangent spaces of any two points of the
manifold and hence that we may take some orthonormal, but otherwise
arbitrary reference basis, which we will call $\vec{u}_a$, and refer
all vectors at all points to that basis (of course, we may identify
$\vec{u}_a$ with its preimage at any given point). Parallel transport
of vectors and tensors can then be introduced as meaning to transport
them keeping constant their components with respect to this basis.
Then Cartan covariant derivative is just the variation with respect to
this reference basis expressed, for example, in terms of the
coordinate basis. Mathematically, this has the consequence of
accepting only parallelizable manifolds as physically meaningful. This
is nothing more than just a topological condition on the sort of
Riemannian spaces we deal with. In particular the coordinate vectors
$\partial_\mu(x)$, no matter which sets of coordinates $x$ we take,
should be expressible in terms of the reference basis:
\begin{equation} 
\partial_\mu(x) = h^{a}{}_{\mu}(x)\vec{u}_a
\end{equation}
Let us also remember that the relationship between the Levi-Civita
covariant derivative $\old{\nabla}$ due to the symmetric Riemannian
connection and the ``Cartan'' covariant derivative $\nabla$ of the
Weitzenb\"ock connection is given by the difference between the
Christoffel symbols of both covariant derivatives:
\begin{equation}
\Gamma^\rho{}_{\mu\nu}=\old{\Gamma}^\rho{}_{\mu\nu}+ K^\rho{}_{\mu\nu}
\end{equation}
 Where $K^\rho{}_{\mu\nu}$ is the contorsion tensor given by:
\begin{equation}
K^\rho{}_{\mu\nu}= \frac{1}{2}\left(g^{\rho\alpha}\left[T_{\mu\alpha\nu} + T_{\nu\alpha\mu}\right] - T^\rho{}_{\mu\nu}\right)
\end{equation}
and hereafter (within the sign conventions adopted in this paper)
$T^\rho{}_{\mu\nu}=\Gamma^\rho{}_{\nu\mu}-\Gamma^\rho{}_{\mu\nu}$ is the torsion
tensor of the Weitzenb\"ock connection. Riemannian connection has
curvature and no torsion and Weitzenb\"ock connection has torsion and
no curvature.  Now, as a first point in the discussion we must clarify
what can be considered a conservation equation for the
energy-momentum tensor within Teleparallel theory.  Suppose we had
some (symmetric) energy-momentum tensor $S_{\mu}^{\;\;\nu}$ and let
$w^\mu$ be the components of some vector which is Cartan covariant
constant. Such vectors do exist because any linear constant
combination of vectors of the reference basis $\vec{u}_a$ is Cartan
covariant constant. It is only when we express them in terms of the
coordinate reference base $\partial_\mu(x)$ that they seem to be
position-dependent. $S_{\mu}^{\;\;\nu} w^\mu$ represents the flow of
the component of energy-momentum in the direction of the four-vector
$\vec{w}$, so $\old{\nabla}_\nu(S_{\mu}^{\;\;\nu} w^\mu)=0$ expresses
the conservation of such $\vec{w}$ component. However we can write
this as:
\begin{eqnarray}\nonumber
0 & = &\old{\nabla}_\nu(S_{\mu}^{\;\;\nu} w^\mu)=\nabla_\nu(S_{\mu}^{\;\;\nu}
w^\mu) - S_{\mu}^{\;\;\rho} w^\mu K^\nu{}_{\rho\nu}\\
& = & w^\mu\left(\nabla_\nu S_{\mu}^{\;\;\nu} - S_{\mu}^{\;\;\rho} T^\nu{}_{\nu\rho}\right)
\end{eqnarray}
If we want conservation of energy-momentum in all directions, then it
must hold:
\begin{equation}
\nabla_\nu S_{\mu}^{\;\;\nu} - S_{\mu}^{\;\;\rho} T^\nu{}_{\nu\rho}=0\label{conserved1}
\end{equation}
To the best of my knowledge this condition has never been put in such a
explicit covariant form by any another author.  This condition is not
the same as the condition $\old{\nabla}_\nu S_{\mu}^{\;\;\nu}=0$. If
we substitute the Cartan covariant derivative by its classical
counterpart, we reach another expression for this conservation law:
\begin{equation}
\old{\nabla}_\nu S_{\mu}^{\;\;\nu} - S_{\sigma}^{\;\;\rho}
K^\sigma{}_{\mu\rho}=0
\end{equation} 
Of course, these equations reduce to the zero divergence condition in
Minkowski space, however the important point is that (for second rank
tensors in Teleparallel spaces) they represent the correct
generalization of the zero divergence condition of Minkowski space.

There is a second way of obtaining the same result which might be
somewhat more transparent and also more easily generalizable to other
types of tensors. Suppose we are given some tensor $S^{\mu\nu}$ and
then we express it partially in terms of the holonomic base
$\partial_\alpha$ and partially in terms of the arbitrary reference
base $\vec{u}_a$. A vector $\vec{V}$ can be expressed in terms of
either base: $\vec{V}= V^\alpha\partial_\alpha=V^\alpha
h^{a}{}_{\alpha}\vec{u}_a= V^a\vec{u}_a$, hence $V^a=V^\alpha
h^{a}{}_{\alpha}$. So let us form the quantities:
\begin{equation}
S^{\mu a}= S^{\mu\nu}h^{a}{}_{\nu}
\end{equation}
As long as we keep fixed the arbitrary reference base, this quantities
will only be transformed in their first index in any coordinate
transformation. And they will be transformed as vectors because the
second index is not transformed. We will just have a set of four
vectors instead of a tensor.  Hence we can apply to them the
divergence theorem, which says (\cite{Lee97}, pg. 43):
\begin{equation}
{\textstyle\int}_M \mbox{\rm div}\vec{V}dv = {\textstyle\int}_{\partial M} <\vec{V},\vec{N}> d\tilde{v}
\end{equation}
Where $\vec{N}$ is just the normal, oriented to the exterior, of
$\partial M$. So to get a conservation equation we just must impose
the condition $\mbox{\rm div}\vec{V}=0$. A expression for this
condition in coordinates is (\cite{Lee03} pg. 384):
\begin{equation}
\mbox{\rm div}\left(V^\alpha\frac{\partial}{\partial x^\alpha}\right) = \frac{1}{\sqrt{-g}}\frac{\partial}{\partial x^\alpha}\left(V^\alpha\sqrt{-g}\right) = 0
\end{equation}
In our case we have the vectors $S^{\mu a}$, so the conservation
equations can be written as:
\begin{equation}
\frac{1}{\sqrt{-g}}\frac{\partial}{\partial x^\mu}\left(S^{\mu a}\sqrt{-g}\right) = 0
\end{equation}
One equation for each component with respect to the reference
base. Now we know that:
\begin{equation}
\frac{1}{\sqrt{-g}}\frac{\partial\sqrt{-g}}{\partial x^\mu} = \old{\Gamma}^\beta{}_{\beta \mu} = \Gamma^\beta{}_{\beta \mu}
\end{equation}
So we can write the conservation equation as:
\begin{equation}
0 = \frac{\partial S^{\mu a}}{\partial x^\mu} + S^{\mu a}\Gamma^\beta{}_{\beta \mu}
\end{equation}
Now let us multiply by $h_a{}^\gamma$ (the inverse matrix of
$h^{a}{}_{\alpha}$), taking into account that:
\begin{equation}
S^{\mu a}=S^{\mu\beta} h^a{}_\beta\quad\Longleftrightarrow\quad S^{\mu a}h_a{}^\gamma = S^{\mu\gamma}
\end{equation}
So we have:
\begin{equation}
0 = \frac{\partial S^{\mu\beta}}{\partial x^\mu} h^a{}_\beta h_a{}^\gamma{} +  S^{\mu\beta}h_a{}^\gamma\frac{\partial h^a{}_\beta}{\partial x^\mu} + S^{\mu\gamma}\Gamma^\beta{}_{\beta \mu}
\end{equation}
Remembering also that:
\begin{equation}
h_a{}^\gamma\frac{\partial h^a{}_\beta}{\partial x^\mu} = \Gamma^\gamma{}_{\beta\mu}
\end{equation}
We reach the conclusion that the conservation equation can be written as:
\begin{equation}
0 = \frac{\partial S^{\mu\gamma}}{\partial x^\mu} + S^{\mu\beta}\Gamma^\gamma{}_{\beta\mu} +  S^{\mu\gamma}\Gamma^\beta{}_{\beta \mu}
\end{equation}
However this is not a manifestly covariant equation, but we can put it
a more convenient form taking into account the definition of the
covariant derivative, written in the form:
\begin{equation}
\frac{\partial S^{\mu\gamma}}{\partial x^\mu}  =  \nabla_\mu S^{\mu\gamma} - S^{\mu\beta}\Gamma^\gamma{}_{\beta\mu} - S^{\beta\gamma}\Gamma^\mu{}_{\beta \mu}
\end{equation}
So substituting we reach the conclusion that the conservation equation
can be written in a manifestly covariant form:
\begin{equation}
0 = \nabla_\alpha S^{\mu\gamma}  + S^{\beta\gamma}T^\mu{}_{\beta \mu}\label{cons2}
\end{equation}
Which is the result we found before so, in some sense, nothing new has
been invented: what has been done is just to rewrite for
parallelizable spaces the condition imposed by the divergence theorem
for having conserved quantities. But now, this also gives a method for
writing conservation equations for higher order tensors. Let
$S^{\alpha\beta\gamma}$ be a third order tensor. By the same method it
is easy to find that the conservation equation can be written as:
\begin{equation}
0 = \nabla_\alpha S^{\alpha\lambda\mu} + S^{\beta \lambda\mu}T^\alpha{}_{\beta \alpha}\label{cons3}
\end{equation}
Very similar to previous equation (\ref{cons2}): just one more
index. The main reason why it is so easy to formulate covariant
conservation equations in these spaces is that we can add magnitudes
which belong to different points, because we can refer all of them to
a common reference (otherwise arbitrary) base $\vec{u}_a$. In a
general Riemannian space, it is only possible add magnitudes defined
in different points if they are scalars, if they are vectors or higher
order tensors, the comparison between them depends on the path used to
carry them from one point to the other, so talking about global values
makes very little sense, unless the magnitude is a scalar. So it is
natural to have conservation equations for scalars (let us say
electric charge), but it is difficult to even think about what is
meant for a conservation equation of a global magnitude which is not a
scalar: it must be clearly stated the coordinate system and how to add
values. And certainly energy or momentum are not scalars.

So there is another reason for considering more seriously
parallelizable spaces as the sort of abstract mathematical spaces to
be used in physical theories: not only they are the only kind of
spaces in which spinorial structures can be defined as was shown in
\cite{Geroch68}, but they are also needed to formulate conservation
equations of magnitudes which are not scalars.

\section{Energy-momentum conservation.}
As it is well known Einstein's tensor $G^{\mu\nu}$ verifies
$\old{\nabla}_\nu G^{\mu\nu}=0$, so from Einstein's equation we do not
get a conservation equation for the energy-momentum tensor of matter
alone.  For the energy-momentum tensor $\cal T $ of matter, accepting
it to be a symmetric tensor, we rather get that $\old{\nabla}_\nu
{\cal T}_{\mu}^{\;\;\nu} = 0$ implies that:
\begin{equation}
 \left(\nabla_\nu {\cal T}_{\mu}^{\;\;\nu} -
{\cal T}_{\mu}^{\;\;\rho} T^\nu{}_{\nu\rho}\right) + {\cal
  T}_{\sigma}^{\;\;\nu}T^\sigma{}_{\nu\mu}  =0 \label{divergmat}
\end{equation}
Comparing it with (\ref{conserved1}) we see it is not quite exactly
the same, a further term is present. Let us remember that we are
dealing with a lagrangian density $\Lambda$ for the gravitational
field which, in Teleparallel theories, might in general be written as
(we take a system of units in which $c=1$):
\begin{equation}
 \Lambda= \kappa_g(a_1\Lambda_1 + a_2\Lambda_2 + a_3\Lambda_3)
\end{equation} 
where:
\begin{eqnarray}
\Lambda_1  & = &  g^{\lambda\mu}T^\alpha{}_{\alpha \lambda}T^\beta{}_{\beta \mu}\qquad
\Lambda_2  =  g^{\lambda \mu}T^\alpha{}_{\beta \lambda}T^\beta{}_{\alpha \mu} \\
\Lambda_3 & = &  T^{\rho \beta \mu}T_{\rho \beta \mu}\qquad
\kappa_g  =  \frac{1}{16\pi G}
\end{eqnarray}
The ``default'' values: $a_1=1$, $a_2=-1/2$, $a_3=-1/4$, produce a
lagrangian density equivalent to General Relativity, meaning that for
those values the difference between $\old{R}$ and $(a_1\Lambda_1 +
a_2\Lambda_2 + a_3\Lambda_3)$ is just a total divergence. As a matter
of fact, adding
$2g^{\alpha\beta}\old{\nabla}_{\alpha}T^{\lambda}{}_{\beta\lambda}$
one obtains $\old{R}$. However, for the moment, we are not going to
fix the value of those coefficients, we want to keep open the
possibility of choosing other values for them.

One way of obtaining the gravitational energy-momen\-tum tensor is to
directly reproduce the reasoning which in classical mechanics leads to
the conservation of energy. Taking $h$ as the determinant of
$h^a_{\;\;\nu}$ we then write (we follow the standard work
\cite{Landau73}, which as a matter of fact is just the first step of
applying Noether's method considering the translational invariance of
the lagrangian, see also \cite{Leclerc2006}, section 2):
\begin{equation}
\frac{\partial (\Lambda h)}{\partial x^\mu}
= \frac{\partial (\Lambda h)}{\partial h^a_{\;\;\nu}}\frac{\partial
  h^a_{\;\;\nu}}{\partial x^\mu} + \frac{\partial (\Lambda h)}{\partial
  h^a_{\;\;\nu ,\mu}}\frac{\partial h^a_{\;\;\nu ,\mu}}{\partial x^\mu}
\end{equation}
The field equations can be written as:
\begin{equation}
\frac{1}{h}h^a_{\;\;\sigma}\left[\frac{\partial (\Lambda h)}{\partial h^a_{\;\;\nu}} -
\frac{\partial}{\partial x^\gamma}\left( h \frac{\partial
\Lambda}{\partial h^a_{\;\;\nu ,\gamma}}\right)\right] + {\cal T}_{\sigma}^{\;\;\nu} = 0\label{Euler}
\end{equation}
And taking them into account one inmediately is led to:
\begin{equation}
{\frac{1}{h}}\frac{\partial}{\partial
  x^\gamma}\left[h\left(h^a_{\;\;\nu ,\mu}\frac{\partial \Lambda}{\partial
    h^a_{\;\;\nu ,\gamma}} - \Lambda\delta^\gamma_\mu\right)\right] -
{\cal T}_{\sigma}^{\;\;\nu}\Gamma^{\sigma}_{\;\;\nu \mu}= 0
\end{equation}
One would like to identify the term within the round brackets in last
equation with the energy-momen\-tum tensor, but it is not a tensor, so
the idea is to decompose it into tensorial and non-tensorial terms,
hence we write previous equation as:
\begin{equation}
\frac{1}{h}\frac{\partial}{\partial x^\gamma}\left[h(Q_{\mu}^{\;\;\gamma} + N_{\mu}^{\;\;\gamma})\right] -
{\cal T}_{\sigma}^{\;\;\nu}\Gamma^\sigma_{\;\;\nu \mu} = 0 \label{entangled}
\end{equation}
Where $Q_{\mu}^{\;\;\gamma}$ is the tensorial part and
$N_{\mu}^{\;\;\gamma}$ is a non-tensorial term. Expressing the
partial derivative of the $Q$ tensor as a Cartan covariant derivative
one arrives inmediately to the following equation:
\begin{eqnarray}\nonumber
 0 & = &\nabla_\gamma Q_{\mu}^{\;\;\gamma}  -  Q_{\mu}^{\;\;\gamma} T^\nu{}_{\nu\gamma} +\\
 & &  + \Gamma^\nu{}_{\nu \gamma}N_{\mu}^{\;\;\gamma} 
+  Q_{\sigma}^{\;\;\gamma}\Gamma^\sigma{}_{\mu \gamma} +  \frac{\partial N_{\mu}^{\;\;\gamma}}{\partial x^\gamma} -
{\cal T}_{\sigma}^{\;\;\nu}\Gamma^\sigma{}_{\nu \mu}\label{nontensorial}
\end{eqnarray}
Now the problem is to eliminate the non-tensorial terms of this
equation. Of course, it must be possible to do so, because it is not
possible to have an equalty between entities which transform in
different ways (just put the first two terms at one side and the other
terms at the other side). The needed calculations for eliminating, or
transforming, those non-tensorial terms are a bit cumbersome, but
with some work it can be shown that by taking $Q_{\mu}^{\;\;\gamma}$
as:
\begin{eqnarray}\nonumber
Q_{\mu}^{\;\;\gamma} & = &
\kappa_g\big\{a_1\left[2\left(g^{\nu\gamma}T^\alpha{}_{\alpha
    \nu}T^\beta{}_{\beta \mu} - g^{\nu\beta}T^\alpha{}_{\alpha
    \nu}T^\gamma{}_{\beta \mu}\right)\right] \\ \nonumber
& + &  a_2\left[2\left( g^{\nu\gamma}T^\alpha{}_{\beta
     \nu}T^\beta{}_{\alpha \mu} - g^{\nu\alpha}T^\gamma{}_{\beta
     \nu}T^\beta{}_{\alpha \mu}\right)\right]\\ \nonumber
& + &  a_3\left[4g^{\beta\gamma}g^{\alpha\nu}g_{\rho \tau}T^\rho{}_{\alpha
    \beta}T^\tau{}_{\nu \mu}\right] \\
 & - &  \left.\big[ a_1\Lambda_1 +a_2\Lambda_2 + a_3\Lambda_3\big]\delta^\gamma_\mu\right\}
\end{eqnarray}
one arrives to the conclusion that
\begin{equation}
 \Gamma^\nu{}_{\nu \gamma}N_{\mu}^{\;\;\gamma} 
+  Q_{\sigma}^{\;\;\gamma}\Gamma^\sigma{}_{\mu \gamma} +  \frac{\partial N_{\mu}^{\;\;\gamma}}{\partial x^\gamma} =\Gamma^\sigma{}_{\mu \nu}{\cal
  T}_{\sigma}^{\;\;\nu} \label{resultofappendix}
\end{equation}
The expression for $Q_{\mu}^{\;\;\gamma}$ might seem strange at first
sight, but it happens to be exactly $-j_{\mu}^{\;\;\gamma}$ (the fully
covariant version of the so called gauge current found in
\cite{Andrade2000}, see also \cite{Vargas2003}, \cite{Maluf2005}):
\begin{equation}
\frac{1}{h}h^a_{\;\;\mu}\frac{\partial (\Lambda h)}{\partial h^a_{\;\;\gamma}} = j_{\mu}^{\;\;\gamma} = -Q_{\mu}^{\;\;\gamma} 
\end{equation}
So substituting (\ref{resultofappendix}) into
(\ref{nontensorial}), we are led to the result:
\begin{equation}
\nabla_\gamma j_{\mu}^{\;\;\gamma} - j_{\mu}^{\;\;\gamma} T^\nu{}_{\nu\gamma} + {\cal T}_{\sigma}^{\;\;\nu} T^\sigma{}_{\mu \nu} = 0 \label{diverggrav}
\end{equation}
So taking into account both equations (\ref{diverggrav}) and (\ref{divergmat})
one gets:
\begin{equation}
\nabla_\gamma({\cal T}_{\mu}^{\;\;\gamma} + j_{\mu}^{\;\;\gamma}) - ({\cal T}_{\mu}^{\;\;\gamma} + j_{\mu}^{\;\;\gamma})T^\nu{}_{\gamma\nu} = 0
\end{equation}
Which has exactly the form of equation (\ref{conserved1}) and so it
can be interpreted as just expressing the conservation of total
energy-momentum: the energy-momentum of the material fields ${\cal
  T}_{\mu}^{\;\;\gamma}$ plus the energy-momentum of the
gravitational field $j_{\mu}^{\;\;\gamma}$. It also clarifies the
meaning of the term ${\cal T}_{\sigma}^{\;\;\nu}T^\sigma{}_{\mu
  \nu}$. This term specifies the energy-momentum interchange between
the gravitational field and the material one. It is the term which
prevents energy-momentum of the gravitational field or
energy-momentum of the material field from being conserved separately
by themselves. The interpretation of $j_{\mu}^{\;\;\gamma}$ as the
correct energy-momentum tensor for the gravitational field can be
further underlined if we rewrite the field equations (\ref{Euler})
as:
\begin{equation}
\frac{1}{h}h^a_{\;\;\sigma}\frac{\partial}{\partial x^\gamma}\left( h \frac{\partial
\Lambda}{\partial h^a_{\;\;\nu ,\gamma}}\right)  = j_\sigma^{\;\;\nu} + {\cal T}_{\sigma}^{\;\;\nu}
\end{equation}
and compares it with the equations for the electromagnetic field in
Minkowski space written as:
\begin{equation}
\frac{\partial}{\partial x^\gamma}\left( \frac{\partial
\Lambda_e}{\partial A_{\nu,\gamma}}\right) = -J^\nu
\end{equation}
Informally, it is usually accepted that this last equation says that
currents are the sources of the electromagnetic field. Then, in the
same sense, the previous equation might be interpreted as saying that
energy-momentum of both the material fields and the gravitational
field is the source of the gravitational field.

However, this is not the final word about the tensor we need, because
$j_\sigma^{\;\;\nu}$ has the uncomfortable characteristic that in
general it is not symmetric. But if we accept the default values for
the coefficients $a_1, a_2, a_3$ then there is an easy way out of the
problem: Einstein's equations are a total of ten equations, however
they do not completely determine the metric tensor. Diffeomorphisms
comprise the gauge freedom in General Relativity: any two solutions
which are related by a diffeomorphism represent the same physical
solution. Now, if instead of considering the metric tensor as the
final solution of a gravitational problem, we ask a complete solution
of such a problem to be given by the specification of the sixteen
functions $h^a_{\;\;\sigma}$ which give the coordinate basis vectors
in terms of the arbitrary constant reference basis, then we have some
further freedom, because once the metric tensor is given, we may have
several ``square roots'' $h^a_{\;\;\sigma}$ which produce the same
metric tensor: There are sixteen arbitrary $h^a_{\;\;\sigma}$
functions and only ten independent conditions imposed by the metric
tensor.  We have lots of ``gauge freedom'', so let us use part of that
freedom to decree that the antisymmetric part of the energy-momentum
tensor $j^{\sigma\nu}$ should be zero. This ``gauge condition''
amounts just to a set of six equations, because in a four dimensional
space an antisymmetric second rank tensor has only six independent
components.  So, although it is a crude way of counting degrees of
freedom, we have increased by six the number of unknown functions when
substituting the metric tensor as solution by the $h^a_{\;\;\sigma}$,
but we have also added six additional equations, so we expect not to
have changed the ``gauge freedom''. The gauge condition can be written
as:
\begin{eqnarray}\nonumber
j^{[\gamma\nu]} = 0 & = &\left( g^{\mu\gamma}T^\nu{}_{\beta \mu}-
g^{\mu\nu}T^\gamma{}_{\beta
  \mu}\right)g^{\lambda\beta}T^\alpha{}_{\alpha \lambda}\\ 
& & -\frac{1}{2}\left( g^{\mu\gamma}T^\nu{}_{\beta \lambda}-
g^{\mu\nu}T^\gamma{}_{\beta
  \lambda}\right)T^{\beta\lambda}{}_{\mu}\label{gaugecondition}
\end{eqnarray}
Of course it is a covariant condition: true in one coordinate system
means true in all, so we are not limiting the set of coordinates in
which the theory is formulated: we are not imposing conditions on the
sort of diffeomorphisms which might be used. As a matter of fact, it
has previously been argued that Teleparallel theories may have too
much gauge freedom (see \cite{Kopczynski82}, \cite{Nester88}) and that
they suffer from a problem of non-predictability of torsion. Although
a formal proof should be investigated, we clearly expect this gauge
condition to fix the origin of such problems. At least, the
introduction of this condition invalidates the reasoning supporting
such assertions, because clearly these additional six equations have
not been taken into account when studying the predictability of
torsion. Furthermore: being an algebraic condition on the torsion
tensor, not every boundary condition is acceptable, because the
boundary condition must also obey the gauge condition. It must be
noted that those problems reappear again \cite{Izumi2013} in more
recent works when analyzing dark energy as is done for example in
\cite{Geng2011} and \cite{Geng2012}.

Accepting such a gauge condition, the energy-momen\-tum of the
gravitational field turns out to be symmetric. And furthermore it is
inmediate to check that it has zero trace. Needless to say it is a
perfectly covariant local definition of energy-momentum for the
gravitational field.

There is one further point which merits some comment. Teleparallel
theories have some degree of freedom in the way the coefficients are
chosen. However it is only for the case in which we obtain the
Teleparallel equivalent to General Relativity (TEGR) when the
resulting equations happen to be symmetric (we obtain Einstein's
tensor). So it is only in this case in which we have the freedom to
impose that the energy-momentum tensor of the gravitational field
should be symmetric. So this condition eliminates the rest of
possibilities for the coefficients.

Anyway, we leave this section with the final expression for the
energy-momentum tensor of the gravitational field in the only case
which interests us: for the ``default values'' of the parameters which
make Teleparallel theory almost equivalent General Relativity, in
the sense that Einstein's equations are also obtained:
\begin{eqnarray}\nonumber
j^{\lambda\gamma} & = &\kappa_g\bigg[T^\alpha{}_{\alpha\nu}\left(T^{\gamma\nu\lambda}+T^{\lambda\nu\gamma}\right) -2T^\alpha{}_\alpha{}^{\gamma}T^\beta{}_\beta{}^\lambda \\ \nonumber
& &  + T^\alpha{}_\beta{}^{\gamma}T^\beta{}_\alpha{}^\lambda -\frac{1}{2}\left(T^\gamma{}_\beta{}^{\alpha}T^\beta{}_\alpha{}^\lambda + T^\lambda{}_\beta{}^{\alpha}T^\beta{}_\alpha{}^{\gamma}\right) \\ 
& & \left. + T_{\tau\alpha}{}^{\gamma}T^{\tau\alpha\lambda} +\left(\Lambda_1 -\frac{1}{2}\Lambda_2-\frac{1}{4}\Lambda_3\right)g^{\gamma\lambda}\right]\label{FinalExprEMT}
\end{eqnarray}

\section{The flat universe case.}\label{TheEnergyContent}
Let us write the gravitational energy-momentum tensor in the specially
important case of flat (homogeneous isotropic) universe. The cases of
positive and negative curvature can be analyzed similarly (they are
also parallelizable spaces) although it is more cumbersome, and it is
done elsewhere \cite{Hermida2009}. As a matter of fact, the class of
parallelizable spaces is quite big, as it is shown in \cite{Geroch70}
(taking into account that every noncompact space, on which a spinorial
structure might be defined, is parallelizable \cite{Geroch68}). We use a
``cartesian'' coordinate system with coordinates $x^0=ct, x^1=x,
x^2=y, x^3=z$, and we postulate the following matrix of gravitational
potential vectors (the role played by the coordinate vectors is
similar to that of vector potentials):
\begin{equation} 
h^a_{\;\;\alpha} =\, {\rm diag}\left( 1, a(t), a(t), a(t) \right) 
\end{equation}
Using this potentials, the metric is just the very well known diagonal
metric of flat space:
\begin{equation}
g_{\nu \mu} = {\rm diag}(1,-a^2(t),-a^2(t),-a^2(t))
\end{equation} 
The energy-momentum tensor for such a space
is:
\begin{equation} 
j^0_0  =   -6\kappa_g\left( \frac{\dot{a}(t)}{a(t)}\right)^2\quad
j^1_1=j^2_2=j^3_3  =   2\kappa_g\left( \frac{\dot{a}(t)}{a(t)}\right)^2
\end{equation}
Where the dot signals ordinary differentiation with respect to
$ct$. The first point which deserves attention is that energy density
is negative, so it seems that there is at least a known field whose
energy density takes negative values. Being proportional to the square
of the Hubble parameter, it can be said that it is purely
``kinetical'' in this case: it is proportional to the square of the
speed at which $a(t)$ changes, and the minus sign tells us that
absortion of (positive) energy will decrese this speed, as energy
density becomes less negative. In the other two cases, of positive and
negative curvature, energy density also happens to turn out negative.

It might be argued that experimental evidence disagrees with
gravitational field having negative energy. After all, there is quite
good experimental indirect evidence of the radiation of gravitational
energy from binary pulsars \cite{Kramer2006} \cite{Stairs2003}.  They
lose energy through radiation. Well, this is just a problem of how to
interpret experimental evidence. An outgoing wave carrying energy
outwards can be interpreted as an outgoing wave of negative energy
incoming particles. A situation similar to a waterpool with an outlet
just in the center of its floor, which is suddenly opened: there is an
outgoing wave of incoming particles. The main difference in this
respect is the sign of the energy of the incoming particles. So to
speak, negative energy gravitons are just falling into the potential
well.

It should be almost intuitively obvious: If there are a couple of
pulsars losing energy through gravitational radiation, then they may
even collapse and form a black hole, so in this situation the
gravitational field increases with time. This is not like the case of
two electrical particles atracting each other, they radiate
electromagnetic energy as they approach each other, but the field
decreases because the dipolar moment decreases when they are nearer. Now,
if the gravitational field were a positive energy field, it is not
obvious how could the gravitational field increase if radiation
consisted in the emission of positive energy gravitons. And if one
accepts it to be a negative energy field, then it is very difficult to
accept that its radiation is formed by positive energy particles.

In the case of a dust-filled universe $a(t) = C_d t^{2/3}$, so the
gravitational energy density is proportional to $-t^{-2}$ which, when
multiplied by $a^3\propto t^2$ to take into account the increase in
volume, just gives constant energy (per comoving volume): dust does
not contribute to any variation of energy of the gravitational field,
it does not interchange energy with the gravitational field.

In the case of a universe filled with just radiation the
solution for $a(t)$ is of the form $a(t)=C_r t^{1/2}$. So the
gravitational energy density is also proportional to $-t^{-2}$ (it is the
square of a logarithmic derivative, so no matter the exponent it will
be proportional to $-t^{-2}$), which when multiplied by $a^3(t)$, to
take into account the increase of volume, gives the result that energy
of gravitational field changes as $-t^{-1/2}$, which is an increase
and which is just the rate needed to compensate the rate at which
energy of radiation decreases: $\rho a^4$ is constant, so $\rho a^3$
decreases as $a^{-1}\propto t^{-1/2}$. The absortion of positive
energy from radiation just decreases the rate at which universe
expands. In the dust-filled case, the decrease in speed is just to
compensate the increase in volume, so that total energy is the
same. As energy of light is absorbed by the gravitational field, its
``kinetic'' energy increases (it decreases its ``speed''). A radiation
dominated universe expands at a slower rate ($t^{1/2}$) than a dust
filled one ($t^{2/3}$): absortion of energy decreases its speed.

Even more clear, for the flat universe Einstein's equation can be
used to calculate the energy-momentum tensor of matter:
\begin{eqnarray} 
{\cal T}^0_0 & = &  6\kappa_g\left(\frac{\dot{a}}{a}\right)^2\label{flatdustmatter}\\
{\cal T}^1_1 & = & {\cal T}^2_2 = {\cal T}^3_3 =  2\kappa_g\left(\frac{\dot{a}^2+2\ddot{a}a}{a^2}\right)
\end{eqnarray}
Looking at equation (\ref{flatdustmatter}) we see that the energy
density of the matter fields is just the same (but positive) as the
energy density of the gravitational field, so that total energy is
zero. This agrees with the idea of a zero initial condition for the
universe.

\section{Dark energy and dark matter.}
We have already argued that equation (\ref{conserved1}) expresses the
conservation of energy-momentum, so let us write that equation as:
\begin{equation}
\Diamond_\nu S_{\mu}^{\;\;\nu}\equiv \nabla_\nu S_{\mu}^{\;\;\nu} - S_{\mu}^{\;\;\rho} T^\nu{}_{\nu\rho}=0
\end{equation}
We have also seen in equation (\ref{divergmat}) that Einstein's
equation implies:
\begin{equation}
\Diamond_\nu {\cal T}_{\mu}^{\;\;\nu} = - {\cal T}_{\sigma}^{\;\;\nu}T^\sigma{}_{\nu\mu}\label{nonconserv}
\end{equation}
Where ${\cal T}$ is the energy-momentum tensor of matter fields. These
means that energy-momen\-tum of material fields is not conserved if
the right hand side is different from zero. Let us suppose we are
dealing with a perfect fluid in an isotropic homogeneous flat
universe. The energy-momen\-tum tensor can be written in such case as:
\begin{equation}
{\cal T}_{\mu\nu}={\rm diag}\left(\rho,pa^2,pa^2,pa^2\right)
\end{equation}
Where $\rho=\rho(t)$ is the mass-energy density, $p=p(t)$ the pressure
and the $a^2(t)$ factors come from the metric.  The right hand of equation
(\ref{nonconserv}) can be easily computed and one obtains:
\begin{equation}
\Diamond_\nu {\cal T}_{\mu}^{\;\;\nu} =\left(- 3\frac{\dot{a}}{a}p(t),0,0,0\right)\label{MatterCreation}
\end{equation}
The first thing which stands out is that if $p(t)\neq 0$ then in an
expanding universe, mass-energy of the material field (by itself) is
not conserved. As a matter of fact the temporal component of this
equation is:
\begin{equation}
\dot{\rho} + 3\rho\frac{\dot{a}}{a} = - 3\frac{\dot{a}}{a}p \label{GRunstable}
\end{equation}
Which just happens to be another form of writing the ``classical
conservation equation'':
\begin{equation}
\old{\nabla}_\mu{\cal T}^{\mu \nu} =0 \quad\Longleftrightarrow\quad \dot{\rho} + 3(\rho + p)\frac{\dot{a}}{a} = 0
\end{equation}
We have already seen such a non-conservation behaviour when we considered the
radiation-filled universe before: a positive pressure means that the
gravitational field drains the positive energy from the ``material''
field. As a matter of fact we have also seen the case $p=0$ in the
dust-filled universe and there was no energy interchange. Let us turn
to the third possible case: $p<0$, which according to
(\ref{MatterCreation}) just means matter creation. Suppose there is some
``spontaneous matter emission'' process, then if matter is created
from the gravitational field, pressure must be negative. This idea can
be stated in different terms: if gravitons happen to be unstable
particles, then they might disintegrate themselves into matter
particles (positive energy) and further gravitons. Then
equation (\ref{MatterCreation}) says that this process produces a negative
pressure component which fuels the acceleration of the expansion.

Of course, in such a case we do not know what exactly to put in the
righthand side of (\ref{GRunstable}). If they exist, we have for the
moment no idea about the characteristics of matter generation
processes from gravitational fields. Anyway these matter-emission
processes must nowadays have very low probability of ocurrence because
otherwise they would have already been detected. Let us just put a
small ``function'' $\lambda$ in the righthand side of equation
(\ref{GRunstable}), and we'll try to guess about it later. Considering
all pressure coming from matter-emission processes, so that all
matter is just dust, then we may write:
\begin{equation}
 \lambda = -3p(t) \frac{\dot{a}}{a}\;\Longleftrightarrow\; 3p(t)
 =-\lambda \frac{a}{\dot{a}}
\end{equation}
So although the process may have very low probability of ocurrence, if
$\lambda$ were constant, the negative pressure would increase with the
Hubble time ($T_H=a/\dot{a}$) or in other words, with the expansion of
the universe. It would of course overcome the mass term in the equation
which determines the acceleration of the expansion of the universe
(one of Friedmann's equations):
\begin{equation}
3\frac{\ddot{a}}{a}=-\frac{1}{4\kappa_g} \left[\rho(t) + 3p(t)\right] = -\frac{1}{4\kappa_g} \left[\rho(t)-\lambda \frac{a}{\dot{a}}\right] \label{firstfried}
\end{equation}
From that moment, positive acceleration sets in. We do not need to
have a cosmological constant to explain the acceleration.  So dark
energy in principle could be explained as the (quantum) process of
dissintegration of gravity into matter. Somehow a graviton emits a
material particle and falls into a lower energy state. This can be
taken as a possible solution to the cosmological constant puzzle
\cite{Bass2011}.

In fact we cannot consider this $\lambda$ to be a constant, it may
depend on the strength of the gravitational field. We would need a
quantum theory of gravitation to be able to calculate
$\lambda(t)$. However, we may get an idea of the order of magnitude of
$\lambda$ by considering zero the acceleration, taking
\cite{Beringer2012} the Hubble constant $H_0=\dot{a}/a$ to be 71
(km/s)/Mpc, and taking the density of the universe to be the critical
one $\approx 9.47\times 10^{-27} \mbox{kg}/\mbox{m}^3$. We get
$\lambda\approx 2.18\times
10^{-44}\mbox{kg}/(\mbox{m}^3\mbox{s})$. This should be the order of
magnitude of the rate at which matter is created at the expense of the
gravitational field nowadays. Of course, it says nothing about what
sort of particles are created. But one may have some intuition:
background gravitons have been expanding since the big bang, so now
they are very long wave particles. Considerig their energy to be
$E_g=-h\nu$ and if they are to be responsible of most matter creation,
then maybe they can only be involved in (relatively) low-energy
processes. So the most likely massive particles to be generated in
this sort of processes nowadays would be very low mass ones, for
example, some sort of low-mass sterile neutrino \cite{Canetti2013},
\cite{Abazajian2012} would be an excellent candidate for dark
(non-detectable) matter. And if energy is so scarce, it is not strange
that neutrinos so created are not relativistic. It is interesting to
note that, although with different motivations, growing matter
scenerios have already been proposed \cite{Amendola2007},
\cite{Wetterich2007}.

There is nothing to prevent the gravitational field from falling even
further down in energy levels (as it is the usual objection to
negative energies), only that nowadays the rate is extremely slow. It
does seem that gravitational systems are in fact unstable. Supposedly,
a quantum theory of gravitation should be able to explain this rate.

It must be noticed that we still have another Friedmann equation (with
$k=0$ as we are considering flat space), or just remembering that
total energy density is zero, adding both contributions we have:
\begin{equation}
\rho=6\kappa_g\left(\frac{\dot{a}}{a}\right)^2 \qquad \Longleftrightarrow\qquad \rho = \frac{3H^2}{8\pi G} \label{fried2}
\end{equation}
The material energy density must be equal to the critical density so,
not having dark energy $\Omega_\Lambda$ as a component of density,
dark matter must be a more substantial contribution to total
density. Dark energy used to serve two purposes: contributing to the
total energy density and providing the acceleration, for which we now
have of another possible mechanism.  Substituting the values of $p$
and $\rho$ in equation (\ref{firstfried}), we obtain:
\begin{equation}
3\frac{\ddot{a}}{a}=-\frac{1}{4\kappa_g} \left[6\kappa_g\left(\frac{\dot{a}}{a}\right)^2 -\lambda \frac{a}{\dot{a}}\right]\label{friedintermedia}
\end{equation}
To get any further we must guess the form of $\lambda$. The equation
that defines $\lambda$ in terms of matter creation is:
\begin{equation}
 \lambda = \dot{\rho} + 3\rho\frac{\dot{a}}{a} = \frac{1}{a^3}\frac{d}{dt}(\rho a^3)
\end{equation}
Where $\rho a^3$ is the matter inside a comoving constant volume. One
is tempted to consider that the number of gravitons should be
approximately constant because the dissintegration rate into matter is
really very weak. Unfortunately that might not be the case. Even if we
dismiss the dissintegration rate into matter (the classical dust-only
universe in which there is no interchange of energy between matter and
the gravitational field) the universe still expands. This means that
the frecuency of gravitons should decrease and hence their total
energy (or the gravitational energy within a comoving volume) should
change if their number is constant. So energy conservation forces us
to assume that their number cannot be constant: There must be
processes in which a graviton dissintegrates itself into other
gravitons of lesser frequency (energy conservation imposes that the
frequencies of the outgoing gravitons should add up to the frequency
of the incoming graviton). 

One would like to think that this increase in the wavelength of the
gravitons is what, when averaged over many gravitons, causes the
universe expansion. But the mechanism must surely be more complicated
because, for example, absortion of energy from photons, which should
decrease the wavelength of gravitons and according to this idea help
expanding the universe, however slows down its expansion compared the
classical dust-only universe.

We have very little base to guess the form of the matter creation
function $\lambda$, but we may suppose that the increase of matter
inside a comoving volume is proportional to both the number of
gravitons within that volume and to the energy of those gravitons, so
we guess that we can approximately express the matter creation rate to
be proportional to the total energy of the gravitons involved.  Now
the energy density of the gravitational field (of the gravitons) is
proportional to $(\dot{a}/a)^2$, so the total energy of gravitons
within constant comoving volume must be $a^3(\dot{a}/a)^2$ and hence
the $\lambda$ function becomes:
\begin{equation}
\lambda =\frac{6k_g}{\tau}\frac{1}{a^3}\left(\frac{\dot{a}}{a}\right)^2a^3=\frac{6k_g}{\tau}\left(\frac{\dot{a}}{a}\right)^2
\end{equation}
Where we have introduced the factor $6k_g$ in the expression just for
commodity and we have written the proportionality constant as $\tau$
because this constant happens to be a time. Anyway, going back to
(\ref{friedintermedia}), it leads to:
\begin{equation}
2\frac{\ddot{a}}{a} = \frac{1}{\tau}\frac{\dot{a}}{a} - \left(\frac{\dot{a}}{a}\right)^2\label{secondtry}
\end{equation}
We can write this equation as:
\begin{equation}
2\frac{\ddot{a}}{a} = \frac{1}{T_H}\left(\frac{1}{\tau} - \frac{1}{T_H}\right)
\end{equation}
Where $T_H=a/\dot{a}$ is the Hubble time. If we make the approximation
that the Hubble time is at every instant the age of the universe
(which of course is not true, but just to get an intuition of what's
going on in the equation), $\tau$ equals the moment in which the
acceleration is zero. Before that moment $T_H$ is less than $\tau$ and
the acceleration is negative, after it $T_H$ is bigger and the
acceleration is positive. So if the age of the universe is about
14.000 million years and acceleration was zero about 6000 million
years ago, we can infer the order of magnitude of $\tau\approx 8\times
10^9$years. The inverse is a small proportionality constant. So let us
try to solve equation (\ref{secondtry}) more exactly.  It has no
closed-form general solution (at least none which Maple$^\copyright$ can
find), but we can try to get a Taylor series in terms of $t$. However
if $\tau\to\infty$, the solution must tend to be that of the
dust-only universe: $C_{d}t^{2/3}$, which has no Taylor series around
$t=0$. So it is better to try first a solution of the form
$C_{dd}t^{2/3}y(t)$ to get rid of the $t^{2/3}$ term. Then $y(t)$ can
be consider as a correction to the dust-only case and its value
should be nearly one in the vicinity $t=0$, if $\tau$ is big
enough. Making such substitution one is inmediately led to:
\begin{equation}
2\frac{\ddot{y}}{y}+ \frac{4}{t}\frac{\dot{y}}{y} =\frac{1}{\tau}\left(\frac{2}{3\tau}+\frac{\dot{y}}{y}\right) -\left(\frac{\dot{y}}{y}\right)^2
\end{equation}
And now we try a solution of the form $y=e^{s(t)}$. We need not put
any constant multiplying this function because it is going to
dissapear, as only quotients between $y(t)$ and its derivatives occur
in previous equation (which is also the reason for such a
trial). Substituting we get:
\begin{equation}
2\ddot{s} + 3{\dot{s}}^2 + \frac{4}{t}\dot{s} - \frac{1}{\tau}\dot{s} - \frac{2}{3t\tau}=0
\end{equation}
This equation has neither a general closed form solution, but we can
get its Taylor expansion by trying:
\begin{equation}
s(t) = \sum_{n=1}^\infty S_nt^n
\end{equation}
We do not need a $S_0$ term because it amounts to a multiplicative
constant. Substituting and equating to zero the terms in $1/t$,
constant, and so on, we get:
\begin{equation}
S_1 = \frac{1}{6\tau};\qquad S_2 = \frac{1}{144\tau^2};\quad\cdots
\end{equation}
Given that $\tau\approx 8\times 10^9$ years, to get $S_2t^2 = S_1t$ we
must wait till $24\tau$, or till $192\times 10^9$ years after the big
bang. We will not usually consider the second term, so for
``small'' times our approximate solution is:
\begin{equation}
a(t) = C_{dd}\,t^{2/3}e^{t/6\tau}
\end{equation}
There are a couple of things which is interesting to find. First of
all, matter density is given by:
\begin{equation}
\rho = 6k_g\left(\frac{\dot{a}}{a}\right)^2 = 6k_g\left(\frac{2}{3t}+\frac{1}{6\tau} + \frac{t}{72\tau^2} +\cdots\right)^2
\end{equation}
So it also begins in a state of infinite density, however the final
density is a more delicate subject: this is not a bad spot to remember
that we are basing our development in just an educated guess about the
properties of matter creation, so any sort of predictions for very
long times should be taken with caution.

The time at which acceleration is zero is not exactly $\tau$ in this
model, but it is given by the condition:
\begin{equation}
0=2\frac{\ddot{a}}{a}=\frac{1}{\tau}\frac{\dot{a}}{a}-\left(\frac{\dot{a}}{a}\right)^2\;\Longrightarrow\;\frac{\dot{a}}{a}=\frac{1}{\tau}\;\Longrightarrow\;\frac{2}{3t} + \dot{s} =\frac{1}{\tau}
\end{equation}
Keeping just the first term in the expansion of $\dot{s}$, and calling
$T_a$ the instant in which the acceleration is zero, this equation
leads to:
\begin{equation}
T_a=\frac{4}{5}\tau
\end{equation}
If one keeps also the second term in the expansion of $\dot{s}$ then
one obtains $T_a=0.7896\tau$ but, given the rough value we are using
for $T_a$, it does not make too much sense to try use such
precision. Anyway, it means that, if we take $T_a$ to be
$8\times10^9$years, then a more reasonable value for $\tau$ is
$10^{10}$years.

More interesting is with how much matter must the universe begin
with. If we multiply density by the cube of the scale factor, we get
the mass per unit of comoving volume:
\begin{equation}
\rho a^3 = C_{dd}^3\,6k_ge^{t/2\tau}\left(\frac{4}{9}+\frac{2}{9}\frac{t}{\tau} + (\frac{1}{36}+ \frac{1}{54})\frac{t^2}{\tau^2}+\cdots\right)
\end{equation}
In the limit $t\to 0$, the mass per comoving volume is:
\begin{equation}
\mu(0) = \frac{8}{3}C_{dd}^3\,k_g
\end{equation}
Taking our instant of time to be $t_0=14\times 10^9$ years, and
$\tau=10^{10}$ the current value of the mass per comoving volume is:
\begin{equation}
\mu_0 \approx 10.2C_{dd}^3\,k_g
\end{equation}
Or 3.8 times the original mass per comoving volume, so this certainly
means that dark matter is more than normal matter nowadays, because all the
excess is due to matter creation. Now, the development we have made
must surely fail for sufficiently early times. It cannot be accepted
that the dissintegration rate of gravitons has no more physics behind
it than the law we have written. It is no more than an approximation
forced by ignorance. In particular, for early enough times one should
need a quantum theory of gravitation to predict the correct rate of
dissintegration (and into what do gravitons dissintegrate). So if
these figures have any sense, they must just reflect the amount of
dark matter generated in the matter dominated era, but as we move
further back into the radiation era things should be described in a
quite different unknown way. So, it is not unreasonable to suppose
that the original matter density of this model is already a mixture of
dark matter and normal matter. So last figure amounts to saying that
about 1/3.8=26.3\% of all matter now present already existed at the
beginning of the matter dominated era. If one accepts that nowadays
normal matter is just about 4.9\% of the critical mass, this implies
that at the beginning of matter dominated era, dark matter was 21.4\%
of the matter now existing, or that it was already 4.37 times more
than normal matter. Recombination time does not coincide with the
moment of matter/radiation equalty, but one can look at the
approximations we are using as taking as instantaneous all the
radiation era, so it does not make too much sense to distinguish
between those moments. Hence this result can be qualitatively compared
with the results of CMB analysis, which tell us the composition of the
universe at recombination. According to Planck Collaboration
data \cite{Planck2013}, the barion density is given by
$\Omega_bh^2=0.022$ (well, with more precision depending on which data
sets you use) and the dark matter density is given by
$\Omega_ch^2=0.12$, so the relation between them is
$\Omega_c/\Omega_b=5.45$; one may think that a 4.37 value is not too
bad a result given the crude model we are using.

However, even this ``agreement'' must be taken with lots of caution,
apart from the precision in the data we have used:
\begin{itemize}
\item First and most important: results of CMB analysis are
  model-dependent, so of course it is not really correct to mix
  results for one model with predictions of another. As a matter of
  fact, only the prediction that present mass (per comoving volume) is
  3.8 times more now than at the beginning of matter-dominated era,
  can be taken as independent of CMB analysis, because it is based on
  Hubble constant, and the moment on which acceleration of the
  universe was zero. Values which can be measured independently of CMB
  analysis.
\item And second: we cannot say exactly at what moment are these
  predictions valid, because we have approximated all the
  radiation era as having zero duration.  
\end{itemize}
So it is only the qualitative result of predicting a greater abundance
of dark matter over normal matter what can be argued in favor of the
model. It must be noted that, according to this model, such
composition has changed substantially since recombination, dark matter
should be much more nowadays: about 95.1\%, or 19.4 times more than
normal matter, if one applies the figures from the $\Lambda$CDM model.
So, although this description transmits a different qualitative
story, there are too many conditionals and guesses in these
figures. Also we would need a better determination of $t_0/\tau$
because, appearing in an exponent, its effect is quite pronounced.

Finally, this model poses questions about the inflation era, because
basically nowadays it seems we are viewing graviton dissintegration as
a highly suppressed process. If a graviton dissintegrates into another
graviton and some further particle(s), the energy difference goes to
the other particle(s). It is then not unreasonable to guess that the
process is more likely if both gravitons have energies of the same
order of magnitude, which means that they must be of sufficiently high
energy to create more massive particles. In sufficiently early times
we should expect to have much more energetic gravitons, because the
lengths involved are much more smaller, and hence frecuencies should
be expected to be higher. So it is expected that many other processes
would not be forbidden and graviton dissintegration into matter, which
drives the acceleration of expansion, should be quite a normal
process, creating in principle all sorts of particles. So one would
not be surprised to find another exponential expansion era at
sufficiently early times. So to speak the idea is that, if there is
such a $\tau$, it should be many orders of magnitude smaller early in
the radiation era and also the intuition is that this early
exponential expansion would end when dissintegration processes become
highly suppressed as the frecuency of gravitons decreases.

\section{Conclusions}
If one takes seriously energy-momentum conservation equations in
Teleparallel theory, then it is unavoidable to reach the conclusion that
negative pressure means matter creation from  the gravitational
field. Saying it otherwise, it means that matter can be created at the
expense of an acceleration of the universe expansion. This eliminates
the problem of the cosmological constant, there is no need for it. It
also means that there is much more dark matter than usually thought,
as we do not have the cosmological constant to add up to the universe
density. In the model introduced the amount of dark matter varies with
time, as more dark matter is generated in the expansion (the model
qualitatively predicts the relation between dark matter and normal
matter). Furthermore, if one accepts that gravitons are unstable, one
may have the intuition that graviton dissintegration processes are
nowadays highly suppressed: we can only observe them indirectly
through the acceleration of universe expansion, however those
processes might have been much more frequent in sufficiently early
times generating the inflation era. If such a point of view could be
held, one could even invoke Occam's razor: what could be the point in
postulating some other fields, like an hypothetical inflaton, if one
could be able to do just with gravity?

\end{document}